\documentclass[aps,preprint]{revtex4-1}  
\usepackage{amssymb}
\usepackage{amsfonts}
\usepackage{amsmath}
\usepackage{mathrsfs}
\usepackage{graphicx}
\usepackage{flushend}
\usepackage{float}
\usepackage{subfigure}
\usepackage{booktabs}

\begin{document}

\title{Repeatedly readable state, spontaneous collapse, and quantum/classical boundary}

\author{Xiao-Fu Peng, Yu-Hang Luo, Jiang Zhu, Bang-Hui Hua, Xue-Nan Chen, Dan-Dan Lian, Zi-Wei Chen, Xiang-Song Chen}

\email{For corresponce: cxs@hust.edu.cn. This is a team work, the authors are listed by the date they joined the group: the first (XFP) is the most recent student, and the last (XSC) is the supervisor.}

\affiliation{Hubei Key Laboratory of Gravitational and Quantum Physics, School of Physics, Huazhong University of Science and Technology, Wuhan 430074, China}

\date{\today}

\begin{abstract}
We propose a model to identify the quantum/classical boundary. 
The model introduces a spontaneous collapse of state superposition:  $\frac{d}{dt} \rho_{ij} =-\frac{i}{\hbar}[H,\rho]_{ij}-\rho_{ij}/\tau_{ij}$. Different from other collapse models, the collapsing scale $\tau_{ij}$ here does not contain a universal parameter, but is specified by the two states $| i\rangle $ and $ | j\rangle$: 
If each state is {\em in principle} repeatedly readable (typically by a QND measurement), then $\tau_{ij}$ is the {\em potentially} needed measuring time to discriminate the two states, and the collapse occurs spontaneously {\em without} any actual monitoring.  
Otherwise,  $\tau_{ij}=\infty$, which means no collapse and everlasting superposition. This happens if one state is not repeatedly readable, or if the two states cannot possibly be discriminated in a particular circumstance (for example in the Rabi oscillation). 
Detailed analysis shows that for a ``trapped Schr{\"o}dinger's cat'', the superposition of $|{\rm here} \rangle$ and $| {\rm there} \rangle $ is forbidden if  $E D \gg 4\pi \hbar c$, and allowed if $E D \le 4\pi \hbar c$, where $D$ is the trap separation and $ E$ is the energy gap, which can be estimated with $  M v^2$. 
For example, if $D\sim 10 \mu {\rm m}$ and $v\sim 10^2{\rm m/s}$, then the quantum/classical boundary is at $M\sim 2\times 10^3 {\rm GeV}/c^2$.
The model also constrains a ``free Schr{\"o}dinger's cat'' to display double-slit interference if $p\theta D\ge 8\hbar$, where $p= Mv$, $\theta $ is the angle spanned by the two trajectories, and $D$ is the slit separation.  For  typical values $v\sim 10^3{\rm m/s}$, $\theta \sim 10^{-5}$, and $D\sim 10{\rm \mu m}$, the quantum/classical boundary is at $M\sim 5 {\rm GeV}/c^2$, which, impressively, is just marginal for the performed experiments with Helium.   
In contrast, this model sets no limit on the coherent length of massless photon,  thus the arm of a Michelson interferometer can be arbitrarily long. 
The spontaneous collapse which we propose can occur for an isolated system, and parallels the decoherence induced by interaction with environment. 

\end{abstract}

\maketitle

\section{the quantum measurement problem}

The so-called quantum measurement problem usually refers to the need of stochastically picking out a single component $\left| \psi \right \rangle _i$ from a superposed quantum state $\left| \psi \right \rangle = \sum_i c_i  \left| \psi \right \rangle _i$ after a measurement, with the probability $|c_i|^2$. \cite{Leggett2002} 
For better appreciation of this renowned problem,  we decompose it into two conceptional and philosophical puzzles, and three falsifiable and physical questions. The two conceptional puzzles are:
\begin{enumerate}
\item A superposed quantum state produces a single outcome for each time of measurement.  

\item The outcome of a quantum measurement can be stochastic, while the Schr\"odinger equation which governs the state evolution is exactly deterministic.  
\end{enumerate}
And the three physical questions are: 
\begin{enumerate}
\item Where is the boundary between classical and quantum behaviors? Or, is there such a boundary at all? Put it graphically: how big can be the celebrated Schr\"odinger's Cat?

\item What or who can make the quantum measurement?
What qualifies for an apparatus or observer, which breaks the quantum superposition and picks out a single outcome? 

\item What should be the outcome in a quantum measurement? This is also called the problem of preferred basis in a quantum measurement. 
\end{enumerate}

All these puzzles and questions make great challenges. Usually, the attempts to interpret (or reinterpret) quantum mechanics focus on the conceptional puzzles, but are rather weak towards the physical questions. Notably, the latter can be answered, within the framework of traditional quantum mechanics, by studying the decoherence effect induced by environmental monitoring. \cite{Schlosshauer2019}  Certainly, not all answers from decoherence have been tested experimentally. Moreover, being just a standard {\em effect} of quantum mechanics, decoherence alone cannot possibly solve the conceptional puzzles of quantum mechanics. 

To completely solve the quantum measurement problem, a natural idea is to modify the standard quantum theory, and taking it as certain limit. \cite{Bassi2013} This, however,  turns out to be extremely difficult, for at least two reasons: 1) Quantum mechanics has been so widely and accurately tested that it is hard to make any essential modification without contradicting experiments.  2) There are not really much clear clue towards the three physical questions listed above. Naively, one might think that quantum behavior is correlated with microscopic masses and distances, while classical behavior belongs to heavy objects and large distances. However, in the LIGO detection of gravitational wave, the reflection mirror weighting tens of kilograms has to be treated quantum mechanically; \cite{LIGO} and in quantum communication experiments, photon entanglement can extend to over a thousand kilometers. \cite{Mozi}

In this paper, we work along a different line. We conjecture that  quantumness is not correlated with microscopic scales, and neither is classicality correlated with macroscopic scales. Instead, we associate classicality with certainty and distinguishability, and assign quantum behavior to uncertain and indistinguishable states. Concretely, we propose a model which demands that superposition of two states with certainty would collapse spontaneously, (namely, without being measured or perturbed,) at a rate specified by the {\em potentially} needed time to discriminate the two states. 

We organize the paper as follows: In Sec. II, we define the model. 
In Sections III and IV, we explain in detail the two key elements of our model, namely the repeatedly readable state, and the spontaneous collapse of the superposition of two such states. 
In Sec. V, we present the key predictions of our model: identification of the quantum/classical boundary. This is done by studying the interfering abilities of a Schr\"odinger's cat, in both trapped and free cases, and of the massless photon.  
In Sec. VI, we summarize our results and make some discussion.  

\section{The model}

Our model does not contain any additional free parameter: 
\begin{equation}
\frac{d}{dt} \rho_{ij} =-\frac {i}{\hbar} [H,\rho]_{ij}-\rho_{ij}/\tau_{ij}.
\label{MasterEq}
\end{equation}
Here, $\rho$ is the density matrix, $H$ is the Hamiltonian, $\rho_{ij}=\left \langle i \right | \rho \left | j \right \rangle$, and $\tau_{ij}$ is a collapsing time scale specific to the two states $| i\rangle $ and $ | j\rangle$.  Note that $\tau_{ij}$ is not an adjustable free parameter, nor does it contain any fixed mass or time scale as in most collapse models. \cite{Bassi2013} Given the properties of states $| i\rangle $ and $ | j\rangle$,  $\tau_{ij}$ can be calculated, as we shall illustrate in detail in the next two sections. At the moment, some important remarks are in order: 
\begin{enumerate}
\item Eq. (\ref{MasterEq}) is to replace the Schr\"odinger equation $ i\hbar \partial _t \left | \psi \right \rangle = H \left | \psi \right \rangle$, and act as the fundamental equation that governs the dynamics of the world. Namely, our model departures essentially from the standard quantum mechanics, and the ordinary Schr\"odinger equation is no longer valid in general. Certainly, our model must not contradict any known phenomena. 

\item The collapse here is spontaneous. For an isolated system,  $[H,\rho] =0$, we have
\begin{equation}
\rho_{ij}(t)=\rho_{ij}(0)e^{-t/\tau_{ij}}.
\label{decay}
\end{equation}
Namely, the coherence between the two states $| i\rangle $ and $ | j\rangle$ decays with the time scale $\tau_{ij}$. Since this decay occurs for an isolated system, it essentially means the {\em collapse of state superposition}. As a result of such collapse, we must take the following position: 

\item State vector in a Hilbert space is no longer a competent description of the world. Namely, even the whole universe cannot be in a pure state. Instead, we have to take the density matrix as the fundamental description, and {\em interpret}  a mixed state as a statistical ensemble.   

\item As a consequence of the above point, the model has built-in stochastic element. 
\end{enumerate}

We now work out the details of our model. 

\section{Repeatedly readable state}

\subsection{Classicallity, certainty, and repeatedly readable state}

Although quantum and classical behaviors can be drastically different, the idea of our model rests heavily on the common features of classical and quantum systems. 
A most important character of the classical object is its certainty or robustness. A coin remains a coin, after being looked, checked, handed, or even worn a little bit. A crucial guarantee for classical  certainty is that the measuring probe can be arbitrarily weak, therefore an object can be known and monitored by measurements with unlimited precision. In contrast, the quantum world is inherently constrained by the Heisenberg uncertainty relation.  It is important, however, to keep in mind that a quantum system also allows for some certainty. In fact, quantization, very much alike the popular word ``digitalization", is often a symbol for accuracy and stability. 

Let us recall that the severe conflict between quantum mechanics and everyday experience lies in the fact that quantum mechanics allows anything to superpose, while a classical object with certainty is not seen to superpose. Then, if we assign some equivalence to classicality and certainty, a possible way to reconcile the quantum/classical conflict is to also restrict the superposition of those quantum states with certainty. 

In this paper, we name a quantum state which displays certainty as ``Repeatedly Readable'' state; or in short, R$^2$ state. (Accidentally, R$^2$ is also the symbol for squared distance, a typical classical quantity.) Naturally, the stable output from a R$^2$ state can be called a R$^2$ value. 

In Sec. IV, we are going to discuss the collapse of the superposition of quantum R$^2$ states. Before doing that, we need to learn well about the properties of such R$^2$ states and R$^2$ values. 

A typical R$^2$ state is the eigenstate of certain observables, and the repeatable way of reading such state is by a quantum non-demolishment (QND) measurement. \cite{Braginsky1992} E.g., the momentum eigenstate of a free particle makes a R$^2$ state, and the R$^2$ value is just the measured momentum. However, we are going to see that the notion of R$^2$ state is not restricted to the eigenstate of the aimed observable, and the R$^2$ value is not necessarily an eigenvalue. We now give some concrete examples which are most relevant for our study. 

\subsection{Discriminating two R$^2$ states: the measuring time}

It is fair to say that the famous ``Schr\"odinger's cat'' is the best example for discussing the quantum/classical conflict or boundary. 
The original version of Schr\"odinger, of the $ | {\rm live} \rangle$ and $ | {\rm dead} \rangle$ states and their possible superposition, though being very graphic, is too hard to define accurately in physics (maybe in medicine as well). It is also brutal to play with the fate of a poor cat. So, we turn to another equally good and much more friendly version:  
the $ | {\rm here} \rangle$ and $ | {\rm there} \rangle$ states and their possible superposition. These states can be perfectly defined and even realized in physics. 

A cat can be pretty quick. However, no matter how quick it is, a cat can never be simultaneously $ | {\rm here} \rangle$ and $ | {\rm there} \rangle$, which is a common practice of quantum particles. Let us repeat that the $ | {\rm here} \rangle$ and $ | {\rm there} \rangle$ states of a classical cat are certainly R$^2$ states. Moreover, they are {\em distinguishable} states, since when speaking of $ | {\rm here} \rangle$ and $ | {\rm there} \rangle$ for a cat, one naturally refers to the macroscopic positions. Let us call attention again to a fact that such distinguishable R$^2$ states are actually also common in quantum mechanics. This fact itself does not make any new knowledge. Our emphasis here is to investigate a key parameter for us: the minimal QND measuring time $\tau_{ij}$ which is needed to distinguish two R$^2$ states $ | i\rangle$ and $ | j\rangle$.

\subsubsection{Schr\"odinger's cat: trapped case}

We consider two spatially separated quantum traps $X_1$ and $X_2$. A particle in one trap, either $X_1$ or  $X_2$, is then a realization of the $ | {\rm here} \rangle=| X_1 \rangle$ or $ | {\rm there} \rangle=| X_2 \rangle$ state. 
It is technically possible to arrange a particle simultaneously in both traps, for example, by first loading the particle into one trap, and adiabatically deforming this trap into two separated traps. \cite{Cronin2009} The fate of such superposed $ | {\rm here} \rangle + | {\rm there} \rangle $ state is the topic of our later sections. Here, we focus on the individual $ | {\rm here} \rangle$ or $ | {\rm there} \rangle$ state, and examine how these two states can be repeatedly distinguished. 

This task is not tough, at least conceptually. A Heisenberg microscope would do the job. \cite{Braginsky1992} Suppose that the two traps are identical, and separated by a spatial distance $D$. A particle in the trap has a discrete spectrum. Let $E$ be the energy gap of the lowest two states, a particle in the ground state can be repeatedly monitored with photons of energy $\hbar \omega \ll E$. Then, if  $\frac 12 D  > \lambda  = 2\pi c/\omega$, the two states $ | X_1 \rangle$ and $| X_2 \rangle$ can be distinguished. The needed measuring time is minimally of the order of $2\tau = 4\pi /\omega$. (A factor $2$ occurs here because one needs $\tau$ to send the photon, and another $\tau$ to receive the scattered photon.) It is very illuminating to combine the two constraints $\hbar \omega \ll E$ and $\frac 12 D > \lambda = 2\pi c/\omega$, we have
\begin{equation}
E D\gg 4\pi \hbar c. 
\label{ED}
\end{equation}

In the above example, $ | X_1 \rangle$ and $| X_2 \rangle$ make two distinguishable quantum R$^2$ states. Note that the R$^2$ value here is the spatial location of the particle, while the particle is not in an eigenstate of position.

\subsubsection{Schr\"odinger's cat: free case}

As a second example, we examine whether a free particle can make the distinguishable quantum R$^2$ states. This will be important for checking the capability of a Schr\"odinger's cat to display double-slit interference. This time, we consider the momentum state and momentum measurement, since the position of a free particle is not a good quantity to manipulate.  

We let a particle of mass $M$ and momentum $p=Mv$ to fly freely from an origin $O$ toward a plate with two parallel slits $S_1$ and $S_2$, located symmetrically with respect to $O$.  The distance of $O$ to the plate is $L$. The two slits are separated by a distance $D$, and each slit has a width $d$. For a typical double-slit experiment, one has $d\ll D \ll L$. We now ask whether the two trajectories $OS_1$ and $OS_2$ can make two distinguishable R$^2$ states.

As we just remarked, we will try to distinguish the two trajectories $OS_1$ and $OS_2$ not by measuring the particle's position during the flight, but by measuring its momentum during the flight. The appropriate measuring device now will not be a Heisenberg's microscope, but be a von Neumann's  Doppler speed meter. \cite{Braginsky1992} Unlike in the Heisenberg's microscope, the frequency uncertainty $\Delta \omega$ of the photon for speed measurement needs to be very small, and accordingly, the photon duration $\tau \sim 1/\Delta \omega$ needs to be large, with an upper limit  set by the flight time: $\tau \le \frac 12 L/v$. (Again, the factor $\frac 12$ is due to the need of sending and receiving the photon to complete the measurement.)

We consider measuring the particle's velocity component $v_\perp $ along the plate. The measurement error is: \cite{Braginsky1992}
\begin{equation}
\Delta v_\perp  =\frac c2 \cdot \frac {\Delta \omega }{\omega }= \frac {c}{2\omega \tau}. 
\label{Deltav}
\end{equation}
Thus, to discriminate the two momenta associated with the two trajectories, we must have
\begin{equation}
\Delta v_\perp  \le \frac 12 v\cdot \frac DL  \simeq \frac 12 v\theta , 
\label{vtheta}
\end{equation}
where $\theta$ is the angle spanned by the two trajectories, or by the two splits with respect to the origin.
Eq. (\ref{vtheta}) sets a lower limit on the applied photon frequency: 
\begin{equation}
\omega \ge \frac{2c}{L\theta}\simeq \frac {2c}{D} .
\label{lower}
\end{equation}

Apparently, the particle's momentum or velocity cannot make a R$^2$ state in this way, since such a  measurement does perturb the particle's velocity by an amount 
\begin{equation}
\delta v_\perp= 2\frac {\hbar \omega}{c} \frac 1 M .
\label{deltav}
\end{equation}

However, if this perturbation lies below the measurement resolution, then it is undetectable, and the momentum state is essentially a R$^2$ state. To this end, we require
\begin{equation}
\delta v_\perp \le \frac 12  \Delta v_\perp .
\label{dD}
\end{equation}

Inspecting the expression of $\Delta v_\perp$ in Eq. (\ref{Deltav}) and the expression of $\delta v_\perp$  in Eq. (\ref{deltav}), we see that the constraint in Eq. (\ref{dD}) can indeed be satisfied, which just sets an upper limit on the  applied photon frequency: 
\begin{equation}
\omega \le \left ( \frac{pc^2 }{2\hbar L} \right)^{\frac 12}.
\label{upper}
\end{equation}

Combining the lower and upper limits on $\omega$ in Eqs. (\ref{lower}) and (\ref{upper}),  we have 
\begin{equation}
\frac {2c}{D} 
\le \omega \le 
\left ( \frac{pc^2 }{2\hbar L} \right)^{\frac 12}.
\label{ul}
\end{equation}

Then, the consistency of the two inequalities sets a very simple constraint on the particle and slit parameters:
\begin{equation}
Mv\frac {D}{L} D \simeq 
p\theta D \simeq p_\perp D \ge 8\hbar.
\label{pD}
\end{equation}
This is the condition for the two trajectories $OS_1$ and $OS_2$, or essentially the momentum states associated with them, to make two distinguishable R$^2$ states. For each trajectory, the minor momentum spreading angle associated with the slit width $d$ might still remain indistinguishable, since we 
we typically have $d\ll D$. 

\bigskip

In the last section of this paper, we will comment on more categories of R$^2$ states. To understand better the characteristics of the R$^2$ state, it is also very illuminating to give some counter examples. 

\subsection{Counter examples of R$^2$ state}

\subsubsection{Rabi oscillation}
A most common but important counter example of R$^2$ state is the quantum system in Rabi oscillation.
For an isolated quantum system with the energy eigenstates $ | E_1 \rangle$ and $ | E_2 \rangle$, we will discuss in the last section whether they can make two distinguishable R$^2$ states. Here, however, we explain that if the system is forced into Rabi oscillation, then no R$^2$ state can exist at all. The reason is rather simple: To put the system into Rabi oscillation, one must apply a radiation field with a frequency not too far from the resonating frequency $\omega _{12}=|E_2-E_1|/\hbar$. In the presence of such a time-dependent perturbing field, a possible R$^2$ state can at best be an adiabatic state. Then, to resolve such an adiabatic state, one must launch a probe with a frequency much higher than $\omega _{12}$, and thus with energy much higher than $|E_2-E_1|$. This probe would then easily destroy the quantum system, and make the measurement non-repeatable. This observation will be fairly important for our discussion of collapsing the state superposition in Sev. IV. 

\subsubsection{Flying photon} 

When discussing the R$^2$ state, the photon trajectory calls for a separate attention. The reason is very special: no matter how the different photon trajectories are measured, the time needed to accumulate the measured information so as to make a possible discrimination cannot possibly be shorter than the traveling time of the photon itself. Namely, the state of a flying photon can only be known afterwards. This property will turn out to be crucial for the interfering ability of the photon in our model. 

\section{Spontaneous collapse}

In Sec. II, we briefly introduced our model by Eq. (\ref{MasterEq}), which is yet empty without specifying the time scale $\tau_{ij}$. In Sec. III, we discussed a key notion for our model, the R$^2$ state, and the associated time scale $\tau_{ij}$ for two R$^2$ states. We used the same symbol  $\tau_{ij}$  in Sections II and III (for a reason that will be evident promptly below). It should be noted, however, that the whole analysis and calculation in Sec. III are performed within the framework of standard quantum mechanics.  
It is when plugging the parameter $\tau_{ij}$ of Sec. III into Eq.  (\ref{MasterEq}) that makes the unique conjecture of our model: Superposition of two R$^2$ states $| i\rangle $ and $ | j\rangle$ would collapse spontaneously at a time scale $\tau_{ij}$, which is the needed time to discriminate $| i\rangle $ and $ | j\rangle$ in a repeatable way. 

It is important to clarify the logic of this model: 
\begin{enumerate}
\item What dictates the fate the superposed state  $|\psi \rangle= a|i \rangle +b | j \rangle $ is not the repeatable readability of the whole state $|\psi\rangle$, but of the individual state $|i \rangle$ and $ | j \rangle $. 

\item What we propose is that {\em if} the two states $|i \rangle$ and $ | j \rangle $ are {\em in principle} repeatedly readable, and can {\em in principle} be repeatedly discriminated at a time $\tau_{ij}$, then  $|\psi\rangle$ would collapse {\em spontaneously}, {\em without any actual monitoring}, to a statistical ensemble of  $|i \rangle$ and $ | j \rangle $.
\end{enumerate}

Whenever a new theory is proposed, its relation to the old theory must be addressed. Unlike many collapse models which take the Schr{\"o}dinger equation as an {\em approximation}, our model agrees {\em exactly} with  the Schr{\"o}dinger equation for the quantum states which are not distinguishable R$^2$ states. For these states, $\tau_{ij}=\infty$, the last term in Eq. (\ref{MasterEq}) drops out, and our model returns to the standard quantum mechanics. We supplement here a definition that the diagonal component $\tau_{ii}=\infty$, which is consistent with the picture that as two states get infinitely close,  $\tau_{ij}$ would become infinitely long.

At the beginning of Sec. III, we briefly mentioned how we came to the idea of our model. Let us say more about it. Our observation and thinking are roughly as follows: 1) If before the measurement a system has to be assigned in superposition, and after the measurement the superposition has to disappear, we must formulated some kind of ``collapse'' or ``reduction''.  2) If we accept the existence of collapse anyhow, and since we definitely desire to give a quantum description to the ``classical'' objects (like a cat) which are not observed to superpose, it is reasonable to conjecture that certain quantum states (including those of the seemingly classical objects) are indeed unable to superpose. 3) A clue to work out this idea is to figure out the essential character of classical objects in a quantum description, and assign such character to the class of quantum states that are unable to superpose. 4) Repeatable readability is the character we identified, which can apply to both classical and quantum systems. This is the idea of R$^2$ state. 5) We cannot brutally forbid the superposition of R$^2$ state states, since such superposition can indeed be created, e.g., in Rabi oscillation. Neither could we let such superposition to collapse instantly, since the Ramsey experiment shows that the superposition can last for some time. 6) Finally, the most natural idea is to associate a finite collapsing time scale, and the good candidate is the discriminating time $\tau_{ij}$, which correlates intimately with the spirit of readability. 

From the above description, one can see that the hint for us comes largely from inspecting the property of states which {\em must collapse}.  However, when conceiving our model we did have also an important hint from inspecting the property of states which {\em must superpose}. This is reflected in the above item 5). Let us dwell a little more about it: 
When planting collapse as a necessary ingredient into a quantum world, one has to take good care of the inherent superpositions.  The observation in item 5) is that the repeatable readability of a state depends on the particular condition. If repeatable readability is spoiled, then collapse stops, and superposition can re-emerge. This is exactly what happens when applying a near-resonance field to create the Rabi oscillation. 

\bigskip 
Before applying our model, we still need to make two important comparisons:

\subsection{Comparison to which-way experiment and complementarity}

The correlation we assigned between collapse and readability may easily be confused with the notion of complementarity in which-way experiment. \cite{Jaeger1995,Englert1996} However, there are critical differences: In the which-way experiment, the path is truly measured, and if the measurement completely identifies the path, then path-superposition collapses immediately and interference disappears. In our model, on the other hand, there is no actual monitoring. If the parameter reaches the constraint in Sec. III.B.2, so that the two paths can {\em in principle} be repeatedly discriminated, then the collapse is an inherent property. Such a collapse is also a relatively mild one, occurring at a finite time scale which one may try to compete to save some interference. However, if the parameters exceed far beyond the constraint, then interference is doomed to fail even without monitoring, no matter how hard efforts one makes.  

Put it another way: While complementarity normally refers to the restriction when {\em measuring} a quantum state, what our model presents might be regarded as a kind of {\em inherent} complementarity when {\em preparing} a quantum state. In our model, one must compromise between distinction and superposition of states: If two states are too distinct, then their superposition might collapse too fast, while closer states are easier to superpose. 

\subsection{Comparison to decoherence}
A pertinent remark here is that decoherence is a ubiquitous quantum effect, while our model is a to-be-tested new quantum theory.  Decoherence is induced by the interaction with the environment, and naturally also finds its place in our model, via the interaction term in the Hamiltonian in Eq. (\ref{MasterEq}). 
Namely, in our model the possible spontaneous collapse parallels the ubiquitous decoherence. They are of totally different mechanisms, and may occur at different rates in a particular circumstance. Like many other collapse models, our model can be tested by experiment only when the predicted collapse happens faster than the traditional decoherence.

Being just a quantum effect, decoherence itself does not introduce the collapse of state superposition. By decoherence, what are suppressed are the interfering terms of the reduced density matrix, obtained by tracing over the unobserved environmental degrees of freedom, while the whole universe is still in a pure state. Therefore, decoherence needs to be combined with certain interpretation of quantum mechanics to address the quantum measurement problem. In comparison, our model makes a complete and reconciled quantum theory (certainly yet to be tested), as we will explain in the final section of this paper. 

Our model does share a common idea with decoherence: entanglement with the environment is crucial to the  behavior of a quantum system. For an entangled system consisting of several subsystems, the collapse in one subsystem would lead to the collapse of the whole entangled state. Analogously, in our model we require that if an entangled system contains a subsystem with distinguishable R$^2$ states, then the whole entangled system would collapse, with the same time scale as inherent to that subsystem. In this way, the orthodox Copenhagen interpretation of quantum mechanics fits perfectly into our model:  All the classical objects in the traditional sense, including the measuring apparatus needed in the Copenhagen interpretation,  are R$^2$ states, and it typically takes very short time to distinguish these states. Therefore, according to our model, as long as an apparatus or observer is involved, the (entangled) state superposition would quickly collapse.

\section{The quantum/classical boundary}

With all the elements of our model explained, we can now explore its prediction for the quantum/classical boundary. The task was essentially done in Sec. III, we just need to associate repeatable readability with classicality. 
We discuss four examples: Schr{\"o}dinger's cat, double-slit interference, photon, and mechanical oscillator.  

\subsection{How big can be a  Schr{\"o}dinger's cat?}
As remarked in Sec. III.B.1, we consider the $|{\rm here} \rangle$ and $|{\rm there} \rangle$ states of a ``trapped cat''.  The quantum/classical criteria is given by Eq. (\ref{ED}): the cat will be classical if $E D \gg 4\pi \hbar c$, and be quantum if $E D \le 4\pi \hbar c$. We can make a quantitative estimation to obtain a sensible figure. To this end, the energy gap $E$ can be roughly taken as $\sim M v^2$.  If we set the typical values $D\sim 10\mu {\rm m}$ and the mean velocity $v\sim 10^2{\rm m/s}$, then the quantum/classical boundary is at a mass $M\sim 2\times 10^3 {\rm GeV}/c^2$. For a much smaller velocity $v\sim 1{\rm m/s}$, which often requires a much lower temperature, the boundary will be at a much heavier mass $M\sim 2\times 10^7 {\rm GeV}/c^2$. In the limit of $v\to 0$, quantum superposition can occur for an infinite mass. (Gravity needs to be ignored, see our remark in the last section.)

Note that in our model the quantum/classical boundary is not a hard one, but a soft one, since the collapse in Eq. (\ref{MasterEq}) takes a finite time which can vary continuously with state parameters. 
We had remarked in Sec. IV that our model can sometimes agree exactly with quantum mechanics. Regarding the example here, this is the case if $E D \le 4\pi \hbar c$, then the two states $|{\rm here} \rangle$ and $|{\rm there} \rangle$ cannot be repeatedly discriminated. In such a case, our model says that the superposition $a|{\rm here} \rangle +b | {\rm there} \rangle $ (of an isolated system) can last forever, and does not spontaneously collapse at all to the individual $|{\rm here} \rangle$ or $|{\rm there} \rangle$ state. 

\subsection{Double-slit interference}
We consider the same set up as in Sec. III.B.2. Then, the quantum/classical criteria is given by Eq. (\ref{pD}).
For a typical double-slit experiment with atoms, \cite{Cronin2009} $v\sim 10^3{\rm m/s}$, $D\sim 10 \mu{\rm m}$, and $\theta \sim 10\mu {\rm m}/{\rm 1m}=10^{-5} $, the quantum/classical boundary is at $M\sim 5 {\rm GeV}/c^2$. Impressively, this is just marginal for the performed experiments with Helium. However, 
as we just remarked earlier, it is better to infer that our model allows for double-slit interference if the mass is below this boundary with the given $v$, $\theta$ and $D$. When the mass goes beyond this boundary, one has to bargain about the detailed time scales to see whether the interference would indeed disappear. For one thing, the collapse given by Eq. (\ref{decay}) takes time; when the parameters barely reach the boundary, the interference just reduces by a factor $1/e$. 
For another thing, the derivations in Sec. III.B.2 related to the Doppler meter is for an object which is ``macroscopic'', namely, with a size much bigger than the wavelength. A mass of $5 {\rm GeV}/c^2$ belongs to the microscopic region, and the Doppler meter would suffer from the angular spreading in photon scattering, and become less efficient. Therefore, for microscopic particles the quantum/classical boundary given by Eq. (\ref{pD}) has to be modified by a factor associated with the differential cross-section of photon scattering on this particle, and the boundary would be shifted towards a larger mass. Without such a detailed investigation, our model can only tell for sure that the interference would definitely disappear for too heavy a mass far beyond the boundary in Eq. (\ref{pD}). 

More importantly, we should keep in mind that the quantum/classical boundary in our model is not for a fixed mass. As we can see clearly from Eq. (\ref{pD}), the particular restriction on mass varies with the velocity $v$, the separation angle $\theta$, and the slit separation $D$ (or equivalently, the traveling distance $L\simeq D/\theta$ or traveling time $L/v$). 

\subsection{How far can the photon interfere?}

The answer in our model is infinite! As we remarked in Sec. III.C.2, monitoring the trajectory of a photon  would take a time which is no less than the traveling time of the photon. Therefore, comparison of photon trajectories could not be finished before completion of the experiment, and the decay term in Eq. (\ref{MasterEq}) makes no sense for the photon. So, we see a very interesting and unexpected mechanism: Relativity can assist the quantum ability! For massless particle like the photon, there is no limit on its coherent length. The arm of a Michelson interferometer can in principle be arbitrarily long. 

It should be remarked, however, that one should hesitate to call the photon an  {\em inherently} quantum particle. As is known in the literature, the energy states of trapped photons can be monitored by QND measurement. According to our model, superposition of such states may suffer a spontaneous collapse. Namely, quantumness/classicality is associated with the particular degrees of freedom of a particle, not with the species of particle in general. 

\subsection{Mechanical oscillator: quantum or classical?}
As our last example, we look at the mechanical oscillator, e.g., the heavy reflecting mirror of LIGO, and check whether our model allows it to behave quantum-mechanically. Suppose that the oscillator has a frequency $\omega_0$ and an amplitude $r_0$ in its ground state. If we try to use Heisenberg's microscope to monitor the coordinate of the oscillator, the minimally meaningful resolution is $\frac 12 r_0$. Then, the probing photon must have a wavelength $\lambda < \frac 12 r_0$, with an energy $E_{\gamma}=2\pi \hbar c/\lambda$ which is far higher than the oscillator energy. Therefore, no matter how heavy the oscillator is, its position in any low-lying state cannot make the R$^2$ state as we define, and our models tells that its position exhibits a quantum behavior.       

Our model can also tell when the oscillator starts to behave classically in position space. If the oscillator is in a state with quantum number $\sim n\gg 1$, the amplitude is roughly $\sqrt n  r_0$. Then, the minimally meaningful spatial resolution becomes $\frac 12 \sqrt n r_0$. By requiring that the energy of the probing photon is much smaller than the oscillator energy $\sim n\hbar \omega_0$, one finds roughly: $n\gg (c/v_0)^{2/3}$, or 
$v_n\sim \sqrt n  v_0\gg v_0 (c/v_0)^{1/3}=c (v_0/c)^{2/3} $, where $v_0$ and $v_n$ are the velocity amplitudes in the ground  and $n$th states, respectively. If these values are allowed by this oscillator, then the oscillator will behave classically. 

\section{Summary and Discussion}

We summarize our work by responding explicitly to the two conceptional puzzles and three physical questions listed at the beginning of this paper. 

Solutions for the two conceptional puzzles of the quantum measurement problem: 
\begin{enumerate}
\item State evolution and quantum measurement are no longer two separate and uncorrelated pieces, they are unified into a single equation. 

\item The world is not described by a state vector, but a density matrix. A mixed state corresponds to a statistical ensemble. Thus, by definition, a single result emerges for each system with a stochastic distribution. 
\end{enumerate}

Solutions for the three physical questions of the quantum measurement problem: 
\begin{enumerate}
\item Yes, there is the quantum/classical boundary, set by repeatable readability and distinguishability.   
Particularly, if two R$^2$ states $| i\rangle$ and $| j\rangle$ can be discriminated in a time $\tau_{ij}$ in a repeatable way, then the superposition of $| i\rangle$ and $| j\rangle$ would spontaneously collapse with the same scale $\tau_{ij}$. 
Since our model does not employ any universal mass or time parameter, the quantum/classical boundary is not universal, but varies from case to case. 

\item The R$^2$ state itself makes the ``observation''. Viewed with our model, the measuring device in the Copenhagen interpretation does nothing but providing some kind of repeatable readability. Namely, the measuring device is just a R$^2$ state. 

\item The R$^2$ state is the ``preferred basis'': if a Hamiltonian allows for R$^2$ states, then such states are the basis for observation. 
\end{enumerate}

The last item even offers a hint on why a time-dependent system also exhibits discrete quantum states. The point is that repeatable readability needs not to be absolute or everlasting. In fact, even the repeatable readability of classical system, like a cat, only lasts for a finite time. Therefore, the instantaneous eigenstate of an adiabatic Hamiltonian can qualify as a (quasi-)R$^2$ state, given that the measuring probe is quick enough to resolve the evolution, while at the same time weak enough so as not to excite the system. 

When computing the collapsing time scale $\tau_{ij}$, we did not intend to be very precise, as our model still appears rather phenomenological. It is not clear what might be the dynamical origin of the spontaneous collapse which we propose, or whether such dynamics exists at all. So, we did not either attempt to arrange our Eq. (\ref{MasterEq}) more elegantly, e.g., into the Lindblad form. \cite{Schlosshauer2019}
Moreover, the key idea in our model, namely spontaneous collapse of distinguishable R$^2$ states, though bearing some physical hints,  is after all fairly bold and radical, thus has to wait for experimental support before making much real sense. 

In this paper, we analyzed the most familiar superposition of states, namely position and momentum states. These are also the cases of steady superposition without energy gap. It is also important to investigate the superposition of the states of other degrees of freedom, e.g., energy and spin. States of these degrees of freedom typically involve energy gap, and thus a superposition has inherently a finite time. It would be very interesting to see whether the energy and spin states can make distinguishable R$^2$ states. If they do, then particle decay would have a different physical picture in our model, as compared to that in the traditional quantum theory. In our model, the decayed particle would automatically lose coherence with the parent particle. Namely, the collapse occurs without a detection. (Recently, a similar picture was proposed by Wilczek with a different model. \cite{Wilczek2021}) In the traditional quantum theory, however, the decayed particle remains coherent with the parent particle, and waits for a detector to collapse the superposition.  

Regarding the experimental investigations of the possible quantum/classical boundary, the most important message brought by our model is that this boundary is not set by a single parameter such as mass, size, or particle number, etc., but set by the correlation of various parameters related to the repeatable readability and distinguishability of the system. Especially, classical behavior is not the privilege of macroscopic objects. Even the electron, proton, and hydrogen atom, which are usually taken as default quantum agents, can be employed to display classical behavior in the sense of collapsing the state superposition. On the other hand, macroscopic objects can possibly exhibit quantum state superposition at microscopic distance and velocity, as suggested by Eqs. (\ref{ED}) and (\ref{pD}). It is misleading to try to find the quantum/classical boundary by simply enlarging the mass, while at the same time reducing the velocity and distance parameters. We would suggest that the microscopic particles such as the electron, proton, and light atoms are actually more advantageous for identifying the quantum/classical boundary, since for these particles the traditional environment-induced decoherence is relatively small compared to that for heavier and larger objects, thus the inherent quantum or classical character of the system can be more cleanly investigated.

To close this paper, we briefly comment on the possibility of monitoring a quantum system by its gravity and charge. If distinguishable R$^2$ states could be identified by such monitoring, then the superposition of quantum states may receive additional constraint from their gravity and charge, which provide long-distance interactions as compared to the contact interaction of photons. These studies are among our future topics. 

We thank Yi-Qiu Ma and Hong-Wei Xiong for very fruitful discussions.  
This work was partly supported by the China NSF via Grant No. 11535005.

\end{document}